\documentclass[onecolumn,showpacs]{revtex4}

\usepackage{graphicx}
\usepackage{dcolumn}
\usepackage{bm}

\input epsf

\begin{document}

\title{\Large Interaction between scalar field and ideal fluid with inhomogeneous equation of state }

\author{\bf Writambhara Chakraborty$^1$\footnote{writam1@yahoo.co.in}
and Ujjal Debnath$^2$\footnote{ujjaldebnath@yahoo.com} }

\affiliation{$^1$Department of Mathematics, New Alipore College,
New Alipore, Kolkata- 700 053, India.\\
$^2$Department of Mathematics, Bengal Engineering and Science
University, Shibpur, Howrah-711 103, India. }

\date{\today}

\begin{abstract}
In this letter we study a model of interaction between the scalar
field and an inhomogeneous ideal fluid. We have considered two
forms of the ideal fluid and a power law expansion for the scale
factor. We have solved the equations for the energy densities.
Also we show that besides being a dark energy model to explain the
cosmic acceleration, this model shows a decaying nature of the
scalar field potential and the interaction parameter.
\end{abstract}


\maketitle

Recent observations of type Ia Supernovae indicate that Universe
is expanding with acceleration [1-5] and lead to the search for a
new type of matter which violates the strong energy condition,
i.e., $\rho+3p<0$. In Einstein's general relativity, an energy
component with large negative pressure has to be introduced in
the total energy density of the Universe in order to explain this
cosmic acceleration. This energy component is known as {\it dark
energy} [6 - 8]. There are many candidates supporting this
behaviour [9], scalar field or quintessence [10] being one of the
most favoured candidates as it has a decaying potential term
which dominates over the kinetic term thus generating enough
pressure to drive acceleration.\\

Presently we live in an epoch where the densities of the dark
energy and the dark matter are comparable. It becomes difficult
to solve this coincidence problem without a suitable interaction.
Generally interacting dark energy models are studied to explain
the cosmic coincidence problem [11, 12]. Also the transition from
matter domination to dark energy domination can be explained
through an appropriate energy exchange rate. Therefore, to obtain
a suitable evolution of the Universe an interaction is assumed and
the decay rate should be proportional to the present value of the
Hubble parameter for good fit to the expansion history of the
Universe as determined by the Supernovae and CMB data [11]. A
variety of interacting dark energy models have been
proposed and studied for this purpose [11-14]. \\

Although a lot of models have been proposed to examine the nature
of the dark energy, it is not known what is the fundamental
nature of the dark energy. Usually models mentioned above are
considered for producing the present day acceleration. Also there
is modified gravity theories where the EOS depends on geometry,
such as Hubble parameter. It is therefore interesting to
investigate models that involve EOS different from the usual
ones, and whether these EOS is able to give rise to cosmological
models meeting the present day dark energy problem. In this
letter, we consider model of interaction between scalar field and
an ideal fluid with inhomogeneos equation of state (EOS), through
a phenomenological interaction which describes the energy flow
between them. Ideal fluids with inhomogeneous EOS were introduced
in [15-17]. Here we have considered two exotic kind of equation
of states which were studied in [18-20] with a linear
inhomogeneous EOS. Here we take the inhomogeneous EOS to be in
polynomial form to generalize the case. Also, the ideal fluid
present here behaves more like dark matter dominated by the
scalar field so that the total energy density and pressure of the
Universe decreases with time. Also the potential corresponding to
the scalar field shows a decaying nature. Here we have considered
a power law expansion of the scale factor, so that we always get
a non-decelerated expansion of the Universe for the power being
greater than or equal to unity. We have solved the energy
densities of both the scalar field and ideal fluid and the
potential of the scalar field. Also a decaying
nature of the interaction parameter is shown.\\

The metric of a spatially flat isotropic and homogeneous Universe
in FRW model is,

\begin{equation}
ds^{2}=dt^{2}-a^{2}(t)\left[dr^{2}+r^{2}(d\theta^{2}+sin^{2}\theta
d\phi^{2})\right]
\end{equation}

where $a(t)$ is the scale factor.\\

The Einstein field equations are (choosing $8\pi G=c=1$)

\begin{equation}
3\frac{\dot{a}^{2}}{a^{2}}=\rho_{tot}
\end{equation}
and
\begin{equation}
6\frac{\ddot{a}}{a}=-(\rho_{tot}+3p_{tot})
\end{equation}

The energy conservation equation ($T_{\mu;\nu}^{\nu}=0$) is
\begin{equation}
\dot{\rho_{tot}}+3\frac{\dot{a}}{a}(\rho_{tot}+p_{tot})=0
\end{equation}

where, $\rho_{tot}$ and $p_{tot}$ are the total energy density and
the pressure of the Universe, given by,
\begin{equation}
\rho_{tot}=\rho_{\phi}+\rho_{d}
\end{equation}
and
\begin{equation}
p_{tot}=p_{\phi}+p_{d}
\end{equation}
with $\rho_{\phi}$ and $p_{\phi}$ are respectively the energy
density and pressure due to the scalar field given by,
\begin{equation}
\rho_{\phi}=\frac{{\dot{\phi}}^{2}}{2}+V(\phi)
\end{equation}
and
\begin{equation}
p_{\phi}=\frac{{\dot{\phi}}^{2}}{2}-V(\phi)
\end{equation}
where, $V(\phi)$ is the relevant potential for the scalar field
$\phi$.\\

Also, $\rho_{d}$ and $p_{d}$ are the energy density and the
pressure corresponding to the ideal fluid with an inhomogeneous
EOS,
\begin{equation}
p_{d}=\omega(t) {\rho}_{d}+{\omega}_{1}f(H,t)
\end{equation}
where, $\omega(t)$ is a function of $t$ and $f(H,t)$ is a function
of $H$ and $t$ ($H$ is the Hubble parameter $=\frac{\dot{a}}{a}$).\\

Now we consider the scalar field interacting with the ideal fluid
with inhomogeneous EOS through an energy exchange between them.
The equations of motion of the scalar field and the ideal fluid
can be written as,
\begin{equation}
\dot{{\rho}_{d}}+3H({\rho}_{d}+p_{d})=-3H{\rho}_{d}\delta
\end{equation}
and
\begin{equation}
\dot{{\rho}_{\phi}}+3H({\rho}_{\phi}+p_{\phi})=3H{\rho}_{d}\delta
\end{equation}
where $\delta$ is a constant.\\

Taking into account the recent cosmological considerations of
variations of fundamental constants, one may start from the case
that the pressure depends on the time $t$ [18]. Unlike the EOS
studied in [19] where the parameters involved in EOS are linear
in $t$, we consider rather a polynomial form. First, we choose the
EOS of the ideal fluid to be,
\begin{equation}
p_{d}=a_{1} t^{-\alpha}{\rho}_{d}-c t^{-\beta}
\end{equation}
where, $a_{1}, c, \alpha, \beta$ are constants.\\

Here, we see that initially the pressure is very large and as
time increases pressure falls down, which is very much compatible
with the recent observational data.\\

We consider a Universe with power law expansion
\begin{equation}
a=t^{n}
\end{equation}
so as to get a non-decelerated expansion for $n \ge 1$, as the
deceleration parameter reduces to $q=-\frac{a\ddot{a}
}{\dot{a}^{2}}=\frac{1-n}{n}<0$.\\

Now equation (10) together with (12) and (13) gives the solution
for ${\rho}_{d}$ to be,

\begin{equation}
{\rho}_{d}=t^{-3n(1+\delta)}e^{\frac{3n a_{1}
t^{-\alpha}}{\alpha}} \left(\frac{3n
a_{1}}{\alpha}\right)^{\frac{3n(1+\delta)+\alpha-\beta}{\alpha}}\frac{c}{a_{1}}
~\Gamma(\frac{\beta-3n(1+\delta)}{\alpha},\frac{3n a_{1}
t^{-\alpha}}{\alpha})
\end{equation}

where, $\Gamma(a,x)$ is upper incomplete Gamma function.\\

Further substitution in the above equations give the solution for
${\rho}_{\phi}, {\dot{\phi}}^{2}$ and $V(\phi)$ to be,
\begin{equation}
{\rho}_{\phi}=3\frac{n^{2}}{t^{2}}-{\rho}_{d}
\end{equation}
\begin{equation}
{\dot{\phi}}^{2}=\frac{2n}{t^{2}}-\left[(1+a_{1}
t^{-\alpha}){\rho}_{d}-c t^{-\beta}\right]
\end{equation}
such that,
\begin{equation}
\phi=\phi_{0}+\int\sqrt{\frac{2n}{t^{2}}-\left[(1+a_{1}
t^{-\alpha}){\rho}_{d}-c t^{-\beta}\right]}~dt
\end{equation}
and
\begin{equation}
 V=\frac{3n^{2}-n}{t^{2}}+\frac{(-1+a_{1}
t^{-\alpha}){\rho}_{d}}{2}-\frac{c t^{-\beta}}{2}
\end{equation}

Since we have considered a power law expansion of the scale
factor so we can see from the above expressions that ${\rho}_{d}$
and $\rho_{\phi}$ are decreasing functions of time so that the
total energy density as well as pressure decreases with time. The
evolution of the Universe therefore can be explained without any
singularity. Normalizing the parameters, we get the variation of
$V(\phi)$ against $\phi$ in figure 3. Equation (18) shows that, for
$\beta<2$, the potential being positive initially, may not retain
this as $t\rightarrow\infty$ (as the 3rd term dominates over first
term and the third and second term being negative for large
values); for $\beta=2$ the potential can be positive depending on
the value of $\left(3n^{2}-n-\frac{c}{2}\right)$ and for,
$\beta>2$ the potential can be either positive depending on the
choices of the constants, but always decreases with time. Hence
$\beta$ is completely arbitrary and depending on various values
of $\beta$ and the other constants, potential to be
positive, although it is always decreasing with time. Fig 3 shows
the nature of the potential for arbitrarily chosen values of the
constants. Also if we consider $w_{d}=\frac{p_{d}}{\rho_{d}},
w_{\phi}=\frac{p_{\phi}}{\rho_{\phi}},
w_{tot}=\frac{p_{tot}}{\rho_{tot}}$, and plot them
(figure 1) against time, we see this represents an XCDM model and
therefore it makes a positive contribution to $\ddot{a}/a$.\\

\begin{figure}
\includegraphics[height=1.7in]{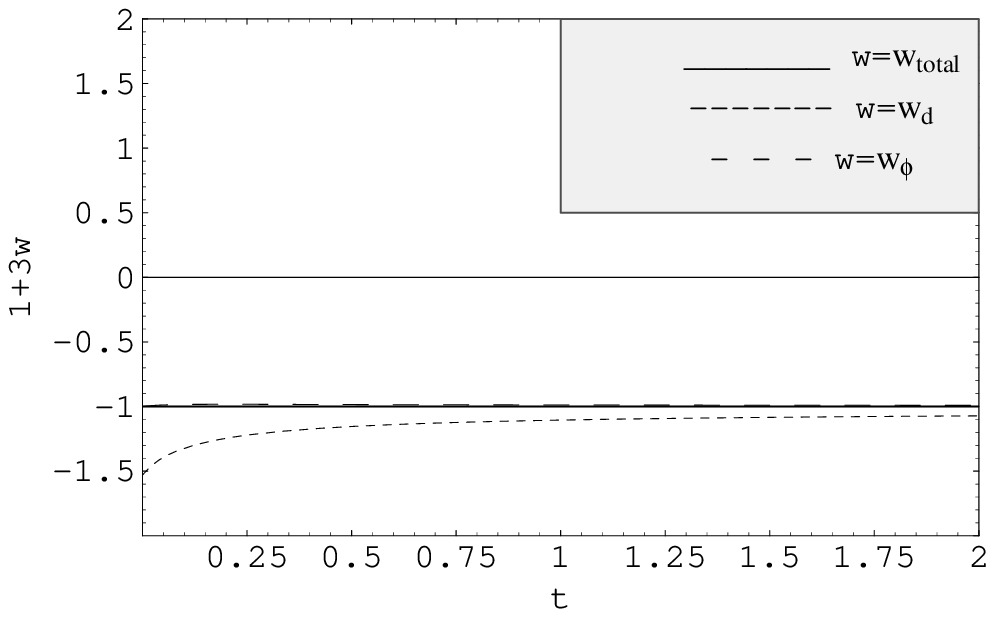}~~~~~~~~~~~~~~~~~~~~
\includegraphics[height=1.7in]{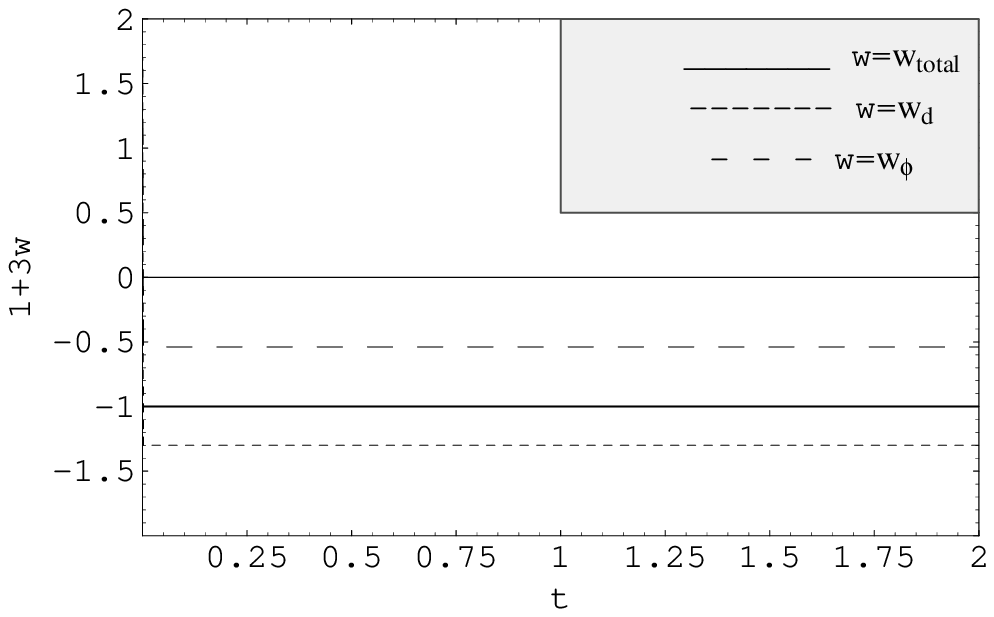}\\
\vspace{1mm}
Fig. 1~~~~~~~~~~~~~~~~~~~~~~~~~~~~~~~~~~~~~~~~~~~~~~~~~~~~~~~~~~~~~~~~~~~~~~~~~~~~~~~~~~~~~~~~~~~~Fig. 2\\

 \vspace{1.8mm}
$n=2,  \alpha=1, \beta=2, a_{1}=.1, c=1, \delta=.01$
~~~~~~~~~~~~~~~~~~~~~~~~~~~~~~~~~~~~~~~~~ $A=\frac{1}{3}, B=-2,
\phi_{0}=1, \delta=.1, n = 2$ \vspace{5mm}

\vspace{5mm} Fig. 1 and 2 show the variation of  $1+3w$ where $w=w_{d},
w_{\phi}, w_{tot}$
 against time normalizing the parameters as mentioned above. \hspace{14cm} \vspace{4mm}

\end{figure}

Inhomogeneous dark energy EOS coming from geometry, for example,
$H$ can yield cosmological models which can avoid shortcomings
coming from coincidence problem and a fine-tuned sudden evolution
of the Universe from the early phase of deceleration driven by
dark matter to the present phase of acceleration driven by dark
energy. Furthermore, such models allow to recover also early
accelerated regimes with the meaning of inflationary behaviors
[20]. The following model is often referred to as Increased
Matter Model where the pressure depends on energy density and $H$.
A detailed discussion of this kind of
EOS can be found in ref. [20].\\

Now we choose the EOS of the ideal fluid to be,
\begin{equation}
p_{d}=A \rho_{d}+B H^{2}
\end{equation}
where, $A$ and $B$ are constants. \\

Considering the power law expansion (13) and using (10) and (19),
we get the solution for $\rho_d$ to be,
\begin{equation}
\rho_{d}=C_{0}
t^{-3n(1+A+\delta)}-\frac{3n^{3}B}{3n(1+A+\delta)-2}t^{-2}
\end{equation}
Further substitution in the related equations yields the solution
for $\rho_{\phi}, \phi, V(\phi)$ to be,
\begin{equation}
\rho_{\phi}=\frac{3n^{2}}{t^{2}}-\rho_{d}
\end{equation}

\begin{equation}
\phi=\phi_{0}+\frac{2}{2-K_{3}}\left[ \sqrt{K_{1}+K_{2}
t^{2-K_{3}}}-\sqrt{K_{1}} \sinh^{-1}
\left(\sqrt{\frac{K_{1}}{K_{2}}} x\right)\right]
\end{equation}
where, $x=t^{\frac{K_{3}}{2}-1}, K_{2}=-C_{0}(1+A),
K_{1}=\frac{6n^{2}(1+A+\delta)-4n-3Bn^{3}\delta+2B n^{2}
}{K_{3}-2}, K_{3}=3n(1+A+\delta)$\\

and
\begin{equation}
V=\frac{3n^{2}-n}{t^{2}}+\frac{A-1}{2} \rho_{d}+\frac{B n^{2}}{2
t^{2}}
\end{equation}

\begin{figure}
\includegraphics[height=1.7in]{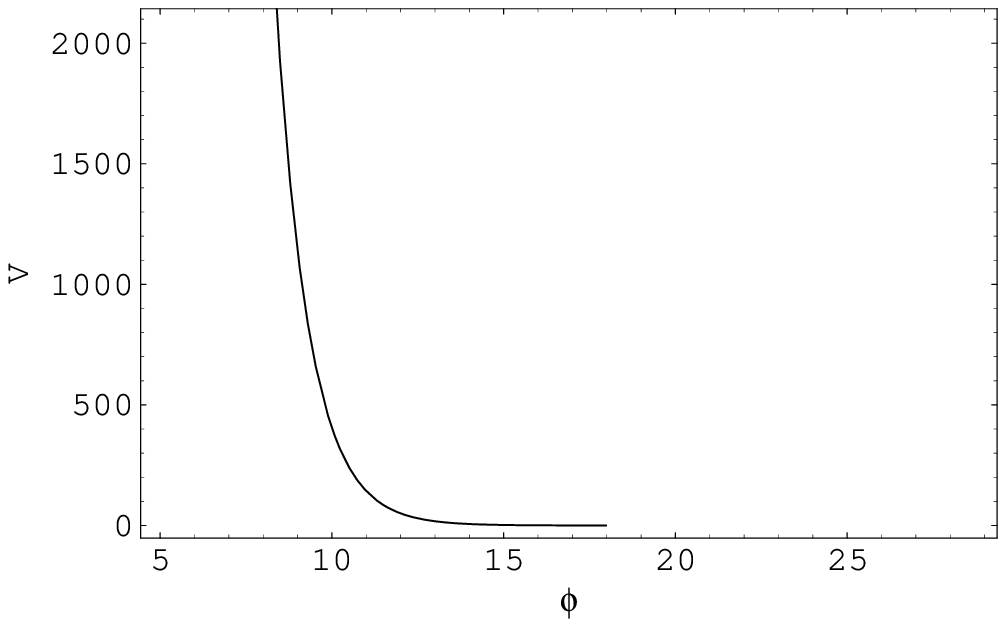}~~~~~~~~~~~~~~~~~~~~
\includegraphics[height=1.7in]{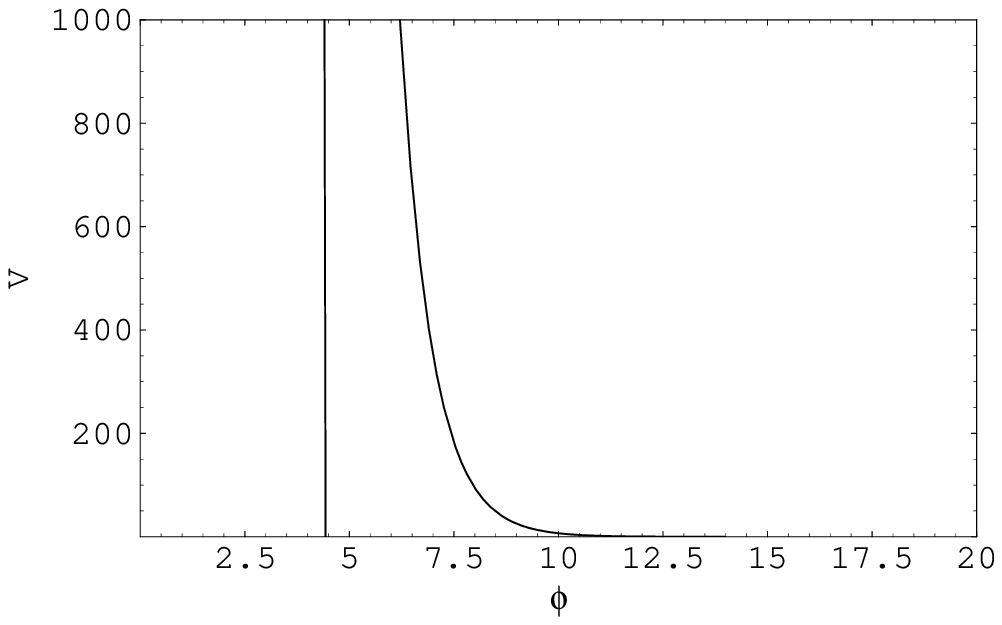}\\
\vspace{1mm}
Fig. 3~~~~~~~~~~~~~~~~~~~~~~~~~~~~~~~~~~~~~~~~~~~~~~~~~~~~~~~~~~~~~~~~~~~~~~~~~~~~~~~~~~~~~~~~~~~~Fig. 4\\

 \vspace{1.8mm}
$n=2,  \alpha=1, \beta=2, a_{1}=.1, c=1, \delta=.01$
~~~~~~~~~~~~~~~~~~~~~~~~~~~~~~~~~~~~~~~~~ $A=\frac{1}{3}, B=-2,
\phi_{0}=1, \delta=.1, n = 2$ \vspace{5mm}

\vspace{5mm} Fig. 3 and 4 show the variation of $V$
 against $\phi$  normalizing the parameters as mentioned above. \hspace{14cm} \vspace{4mm}

\end{figure}

Equation (22) shows that $K_{1}$ must be positive and hence
$K_{2}$ also must be positive for a valid expression. Also
equation (20) says that $C_{0}$ must be positive, otherwise
$\rho_{d}$ becomes negative initially. Therefore expression of
$K_{2}$ says that $A$ must be negative, in fact, $A<-1$, such that
depending on the value of $B$ pressure can be positive or
negative. Normalizing the parameters, we get the variation of $V$
against $\phi$ in figure 4. The figure shows a decaying nature of
the potential. Also if we consider
$w_{d}=\frac{p_{d}}{\rho_{d}},
w_{\phi}=\frac{p_{\phi}}{\rho_{\phi}},
w_{tot}=\frac{p_{tot}}{\rho_{tot}}$, and plot them
(figure 2) against time, like the previous case, we see this
represents an XCDM model and therefore it makes a positive
contribution to $\ddot{a}/a$.\\

In this letter we study a cosmological model of the Universe in
which the scalar field has an interaction with an ideal fluid
with inhomogeneous EOS. The interaction is introduced
phenomenologically by considering  term parameterized by the
product of the Hubble parameter, the energy density of the ideal
fluid and a coupling constant in the equations of motion of the
fluid and the scalar field. This type of phenomenological
interaction term has been investigated in [12]. This describes an
energy flow between the scalar field and the ideal fluid. Also we
consider a power law form of the scale factor $a(t)$ to keep the
recent observational support of cosmic acceleration. For the
first model putting $c=0, \alpha=0$ we get the results for
barotropic fluid. Here for $\alpha$ and $\beta$ to be positive,
the ideal fluid and the scalar field behave as dark energy. Also
we see that the interaction term decreases with time showing
strong interaction at the earlier stage and weak interaction
later. Also the potential corresponding to the scalar field is
positive and shows a decaying nature. In the second model where
$p_{d}$ is a function of $\rho_{d}$ and the Hubble parameter $H$,
we see that the energy density and the pressure of the ideal
fluid and that of the scalar field always decreases with time.
From figures 3 and 4, we see that, the potential function $V$
decreases for both decelerating ($n<1$) and accelerating phase
($n>1$). Also from the values of density and pressure terms, it
can be shown that the individual fluids and their mixtures
satisfy strong energy condition for $n<1$ and violate for $n>1$.
A detailed discussion of the potential of a scalar field can be
found in ref. [21]. We see that the coupling parameter shows a
decaying nature in both the cases implying strong interaction at
the early times and weak interaction later. Thus following the
recipe provided in ref. [22] we can establish a model which can
be a suitable alternative to dark energy explaining the decaying
energy flow between the scalar field and the fluid and giving
rise to a decaying potential. As a scalar field with potential to
drive acceleration is a common practice in cosmology [22], the
potential presented here can reproduce enough acceleration
together with the ideal fluid, thus explaining the evolution of
the Universe. Also we have considered inhomogeneous EOS
interacting with the scalar field which can represent an
alternative to the usual dark energy model. However, stability
analysis and spatial inomogeneity analysis [10] are more complicated
for our investigation, since we are considering the ideal fluid with
two types of equation of states and are analysing whether they can be considered as
an alternative to dark energy. Also we have seen that the
equations of motion (10) together with the given form of the
pressure (12) and (19) are difficult to solve unless we consider
the power law form (13). Once the power law form is considered,
we can easily find exact solution of $\rho_{d}$ (from eq.(10))
and hence $\rho_{\phi}$ (from eq.(2)), which lead to the given
expression for the potential $V(\phi)$ [from eqs. (7), (8)] analytically.
Though this is the backward approach, but otherwise if we start from
$V(\phi)$ i.e., say $V(\phi)=V_{0}Exp(-k \phi)$,
we cannot find any exact solution of $\rho_{d}, \rho_{\phi},
p_{d},p_{\phi}, \phi, a $. So we can only draw conclusions
graphically, not analytically. For example, Ellis et al [23] have
discussed for the model with radiation and scalar field and found exact
solutions in the backward approach. \\

{\bf Acknowledgement:}\\

The authors are thankful to IUCAA, India for warm hospitality
where part of the work was carried out. Also UD is thankful to
UGC, Govt. of India for providing research project grant (No. 32-157/2006(SR)).\\

{\bf References:}\\
\\
$[1]$ S. J. Perlmutter et al, {\it Bull. Am. Astron. Soc.} {\bf
29} 1351 (1997).\\
$[2]$ S. J. Perlmutter et al, {\it Astrophys. J.} {\bf 517} 565
(1999).\\
$[3]$ A. G. Riess et al, {\it Astron. J.} {\bf 116} 1009 (1998).\\
$[4]$ P. Garnavich et al, {\it Astrophys. J.} {\bf 493} L53
(1998).\\
$[5]$ B. P. Schmidt et al, {\it Astrophys. J.} {\bf 507} 46 (1998).\\
$[6]$ V. Sahni and A. A. Starobinsky, {\it Int. J. Mod. Phys. A}
{\bf 9} 373 (2000).\\
$[7]$ P. J. E. Peebles and B. Ratra, {\it Rev. Mod. Phys.} {\bf
75} 559 (2003).\\
$[8]$ T. Padmanabhan, {\it Phys. Rept.} {\bf 380} 235 (2003).\\
$[9]$ E. J. Copeland, M. Sami, S. Tsujikawa, {\it Int. J. Mod.
Phys. D} {\bf  15} 1753 (2006).\\
$[10]$ P. J. E. Peebles and B. Ratra, {\it Astrophys. J. Lett.}
{\bf 325} L17 (1988); B. Ratra and P. J. E. Peebles, {\it Phys.
Rev. D} {\bf 37}
3406 (1988).\\
$[11]$ M. S. Berger, H. Shojaei, {\it Phys. Rev. D} {\bf 74}
043530 (2006).\\
$[12]$ R.-G. Cai, A. Wang, {\it JCAP}
{\bf 03} 002 (2005).\\
$[13]$ W. Zimdahl, {\it Int. J. Mod. Phys. D} {\bf 14}2319
(2005)\\
$[14]$ B. Hu, Y. Ling, {\it Phys. Rev. D} {\bf 73} 123510
(2006).\\
$[15]$ S. Nojiri, S. D. Odintsov, {\it Phys. Rev. D} {\bf 72}
023003 (2005).\\
$[16]$ S. Nojiri, S. D. Odintsov, {\it Phys. Lett. B} {\bf 639}
144 (2006).\\
$[17]$ E. Elizalde, S. Nojiri, S. D. Odintsov, P. Wang, {\it Phys.
Rev. D} {\bf 71} 103504 (2005).\\
$[18]$ I. Brevik, S. Nojiri, S. D. Odintsov, L. Vanzo, {\it Phys.
Rev. D} {\bf 70} 043520 (2004).\\
$[19]$ I. Brevik, O. G. Gorbunova, A. V. Timoshkin, {\it Eur.
Phys. J. C} {\bf 51} 179 (2007).\\
$[20]$ S. Capozziello, V. Cardone, E. Elizalde, S. Nojiri, S. D.
Odintsov, {\it Phys. Rev. D} {\bf 73} 043512 (2006).\\
$[21]$ V. H. Cardenas, S. D. Campo, {\it Phys. Rev. D} {\bf 69}
083508 (2004).\\
$[22]$ T. Padmanabhan, {\it Phys. Rev. D} {\bf 66} 021301
(2002).\\
$[23]$ G. F. R. Ellis, M. S. Madsen {\it Class. Quantum Grav.} {\bf
8} 667 (1991).\\

\end{document}